\documentclass[10pt]{article}
\usepackage{multirow}
\usepackage[dvips]{graphicx}

\usepackage[psamsfonts]{amssymb}
\usepackage{amsxtra}
\usepackage{threeparttable}
\usepackage{amsmath}
\usepackage{amssymb}

\usepackage{graphicx}

\usepackage{cite}

\usepackage{color}

\usepackage{setspace}
\usepackage{float}

\usepackage{url}

\usepackage[labelfont=bf,labelsep=period,justification=raggedright]{caption}

\makeatletter
\renewcommand{\@biblabel}[1]{\quad#1.}
\makeatother

\date{}

\begin{document}

\begin{flushleft}
{\Large
\textbf{Psychophysical Responses Comparison in Spatial Visual, Audiovisual, and Auditory BCI-Spelling Paradigms}
}

Moonjeong Chang$^1$, 
Nozomu Nishikawa$^1$, 
Zhenyu Cai$^1$, Shoji Makino$^1$, 
and Tomasz M. Rutkowski$^{1,2,\ast}$
\\
\bf{1} TARA Life Science Center, University of Tsukuba, Tsukuba, Japan\\
\bf{2} RIKEN Brain Science Institute, Wako-shi, Japan\\
$\ast$ E-mail: tomek@tara.tsukuba.ac.jp
\end{flushleft}

\section*{Abstract}

The paper presents a pilot study conducted with spatial visual, audiovisual and auditory brain-computer-interface (BCI) based speller paradigms. The psychophysical experiments are conducted with healthy subjects in order to evaluate a difficulty and a possible response accuracy variability. We also present preliminary EEG results in offline BCI mode. The obtained results validate a thesis, that spatial auditory only paradigm performs as good as the traditional visual and audiovisual speller BCI tasks.

\section{Introduction}

The brain computer and brain machine interfaces (BCI and BMI)~\cite{bciBOOKwolpaw}, or generally the brain evoked responses to any external stimuli are usually based on the monitoring of the neuro-- and electro--physiological activity by means of the electroencephalogram (EEG)~\cite{book:eeg}. Due the method's non-invasive nature, the EEG based BCI paradigms are perfectly suited to be at the core of future ``intelligent'' interfacings and neuro--prosthetic developments. The concepts are particularly fitted to the needs of the handicapped as well the cores of the smart environments, computer gaming and virtual reality applications.

A concept of utilizing brain various sensory modalities creates a very interesting possibility to let the subjects to interact with more rich multi-sensory environments, which possibly shall result with better brain--wave--responses classification and information--transfer--rates (ITR).

The auditory BCI (aBCI) is thus potentially a less mentally demanding paradigm receiving recently more attention in auditory neuroscience applications~\cite{iwpash2009tomek,tomekAPSIPA2010,bciSPATIALaudio2010,tomekHAID2011,assrBCI2012}. We propose to compare first psychophysical responses to various audio or visual modality reposes in application to a spatial perception typer/speller paradigm. The results shall next help to design a more comfortable BCI speller/typer paradigm.

Next we present the preliminary results from EEG experiments where the evoked responses to the various modality in form of event related potentials (ERP) are compared for target and non--target stimuli.

In the following sections we discuss audio, visual and audiovisual spatial speller BCI paradigm interface applications. Next we describe the psychophysical and EEG offline BCI experiments conducted for auditory, visual and audiovisual cases. A statistical analysis of the obtained results and a discussion conclude the paper.



\section{Methods}

The experiments are conducted in a combination of \textsf{BCI2000} environment~\cite{bci2000book} to generate stimuli patterns and \textsf{Max/MSP} by Cycling'74 software to analyze subjects behavioral responses in form of button presses to instructed target letter directions in visual, auditory and audiovisual modes. For the psychophysical and the later EEG BCI experiments the stimuli in various modalities are generated as follows.

\subsection{The psychophysical experiments}

The psychophysical experiment series is conducted with three subjects seating in front of computer display where instructions and visual stimuli are given. The subjects respond by pressing a button immediately after each target's appearance. The three tested modality setting details are as follows. 

\subsubsection{The visual speller psychophysical task}

Five Japanese hiragana letters \emph{ - a, i, u, e, o -} are flashed in a random order. The subject are instructed to attend (press a response button) to the target letter presented in each random trial sequence. The non--target stimuli shall be ignored.

\subsubsection{The auditory speller psychophysical task}

The same Japanese hiragana letters are delivered as synthetically generated sounds from two loudspeakers positioned in front of the subjects at a one meter distance and $-45^\circ$ and $45^\circ$ azimuths. A vector based amplitude panning method (VBAP)~\cite{pulkkiVBAP1997} is used to virtually distribute sound images at the positions of $-90^\circ, -45^\circ, 0^\circ, 45^\circ, 90^\circ$.
The sound levels are uniform for all the letters at $70$~dB. Here again subjects are instructed to attend (press the response button)  as soon as they could hear a target letter in a randomly presented series.

\subsubsection{The audiovisual speller psychophysical task}

The audiovisual task involves presentation of the both above described auditory and visual stimuli letters simultaneously creating a multimodal paradigm. The instructions given to the subjects are the same as in the both unimodal tasks.

\subsection{The EEG experiment with offline BCI protocol}

We conduct experiments to record EEG for further offline analysis of the brain responses to visual, auditory and audiovisual modalities. The purpose is to find how much the $P300$ EEG potential (a so called ``aha'' response) variability, for target versus non-target discrimination in each modality, is modulated by various sensory stimuli in the same spatial speller task. The EEG signals are captured with eight dry electrodes portable wireless system by \textsf{g.tec} (\textsf{g.SAHARA \& g.MOBIlab+}). The recorded EEG at each electrode is preprocessed by \textsf{BCI2000} application. The sampling rate is $256$~Hz and the notch filter to remove electric power lines interface of $50$~Hz is set in a band of $48-52$~Hz, according to the East Japan power stations specification.
The EEG experiments are conducted with agreement of the institutional ethical committee guidelines for experiments with human subjects.
The EEG experimental protocol is as follows:

\subsubsection{The visual speller offline BCI task}

Five Japanese hiragana letters are flashed in a random order. The subjects are instructed to attend and count the target letters presented in each random trial, while ignoring the non-targets. Since only the EEG evoked response are of the experimental interest, the subjects are requested not to move and to limit eye blinking.

\subsubsection{The auditory speller offline BCI task}

The same Japanese hiragana letters are delivered as synthetic generated sounds from two loudspeakers positioned in front of the subjects at a one meter distance at the $-45^\circ$ and $45^\circ$ azimuths. This experiment is conducted with the conditions as the above visual stimuli EEG recording, thus the subjects receive the same instructions.

\subsubsection{The audiovisual speller  offline BCI task}

The audiovisual task involves presentation of the both auditory and visual stimuli letters as in the previous sections. The subject is given the same instruction to attend and count the audiovisual presented target letter as in above unmoral offline BCI tasks.

\section{Results}

The results of the psychophysical experiments in the three modality spatial speller settings have shown no significant differences among median response delays. The visual, auditory and audiovisual modalities the have similar cognitive loads resulting with the same behavioral markers. There has been no significant median response differences in pairwise \emph{t-} and \emph{Wilcoxon-tests}. The grand mean results for all subjects have been presented in Figures~\ref{fig:audio}--\ref{fig:audiovideo} for auditory, visual and audiovisual modalities respectively. A summary plot comparing all three modalities with median values for all stimuli has been presented in Figure~\ref{fig:multimodal}. A median value of mean psychophysical responses, in the above figure, for auditory modality lies between audiovisual and visual modalities. Here again the differences among median values has been not significant.

\begin{figure}[H]
	\centering
	\includegraphics[width=0.7\linewidth]{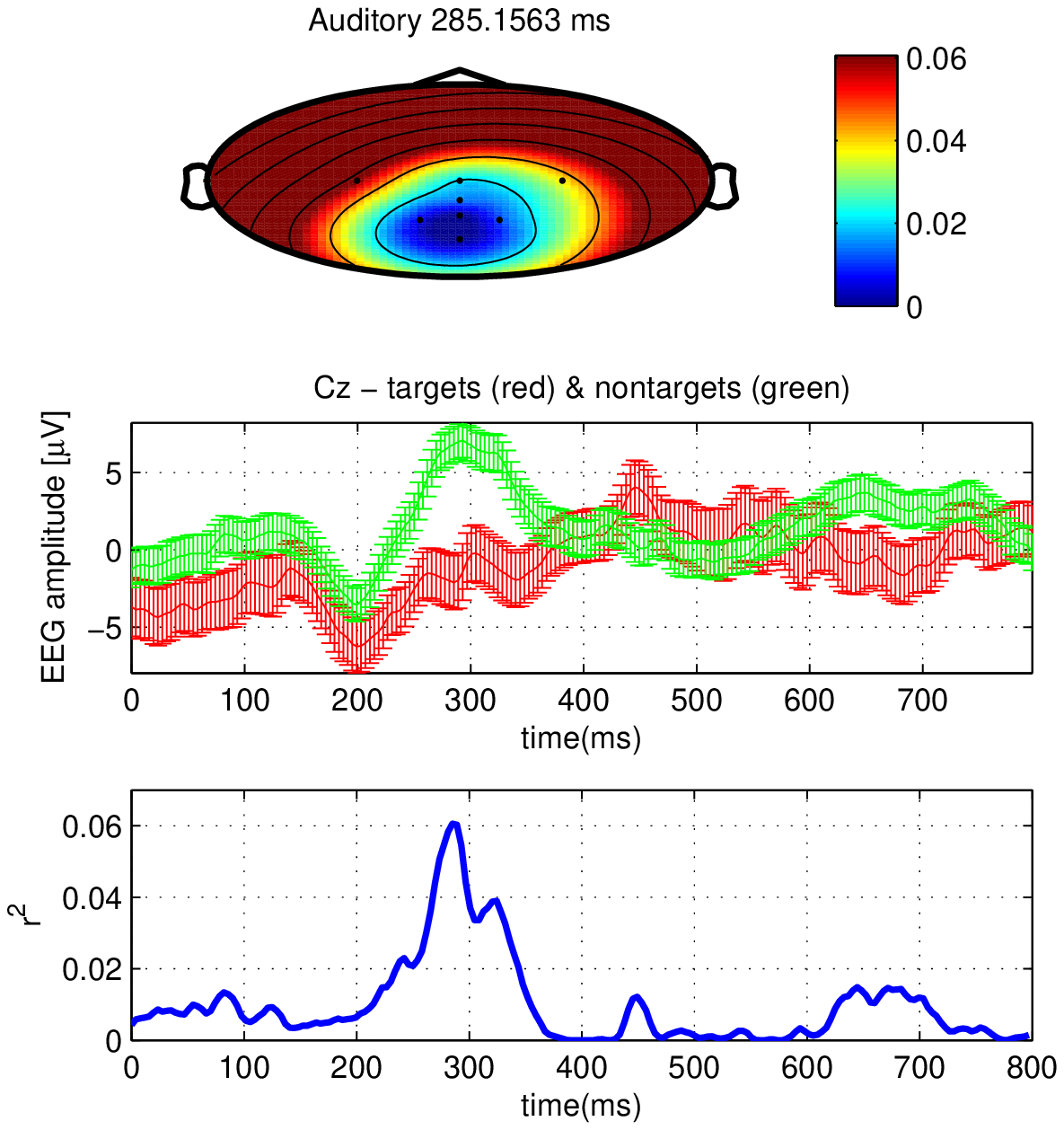}
	\caption{Auditory modality only offline BCI EEG responses presented as scalp mean topography for targets - in the top panel; as the mean time series with standard deviation error--bars for non-targets (green) and targets (red) in the middle panel; and as the signed statistical difference (i.e., signed $r^2$) value which evaluates the discriminability between the two types of ERPs - in the bottom panel. The figure was created with \textsf{P300GUI}~\cite{p300gui}.}\label{fig:erpAUDIO}
\end{figure}

\begin{figure}[H]
	\centering
	\includegraphics[width=0.7\linewidth]{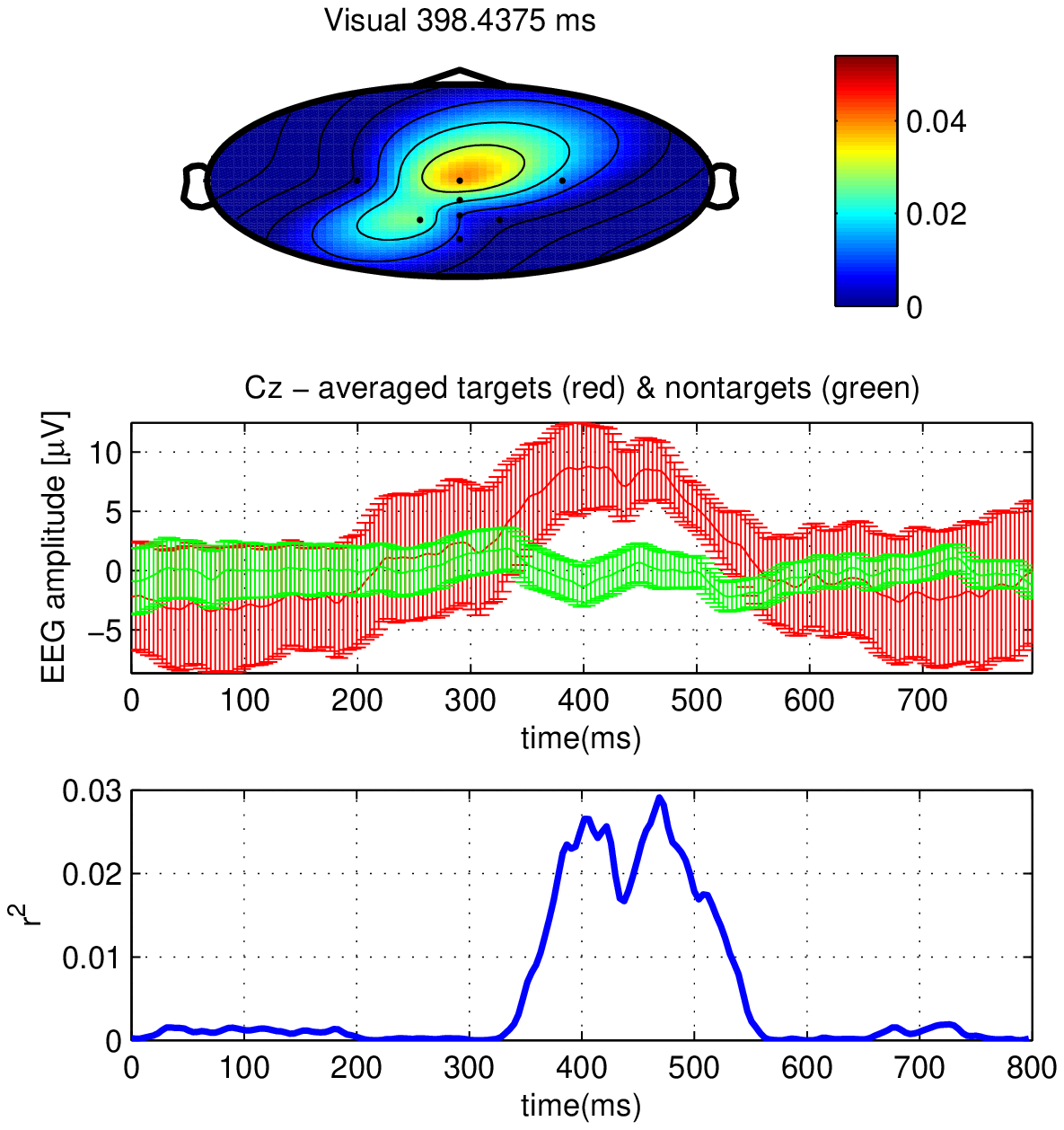}
	\caption{Visual modality only offline BCI EEG responses presented as scalp mean topography for targets - in the top panel; as the mean time series with standard deviation error--bars for non-targets (green) and targets (red) - in the middle panel; and as the signed statistical difference (i.e., signed $r^2$) value which evaluates the discriminability between the two types of ERPs - in the bottom panel. The figure was created with \textsf{P300GUI}~\cite{p300gui}.}\label{fig:erpVISUAL}
\end{figure}

\begin{figure}[H]
	\centering
	\includegraphics[width=0.7\linewidth]{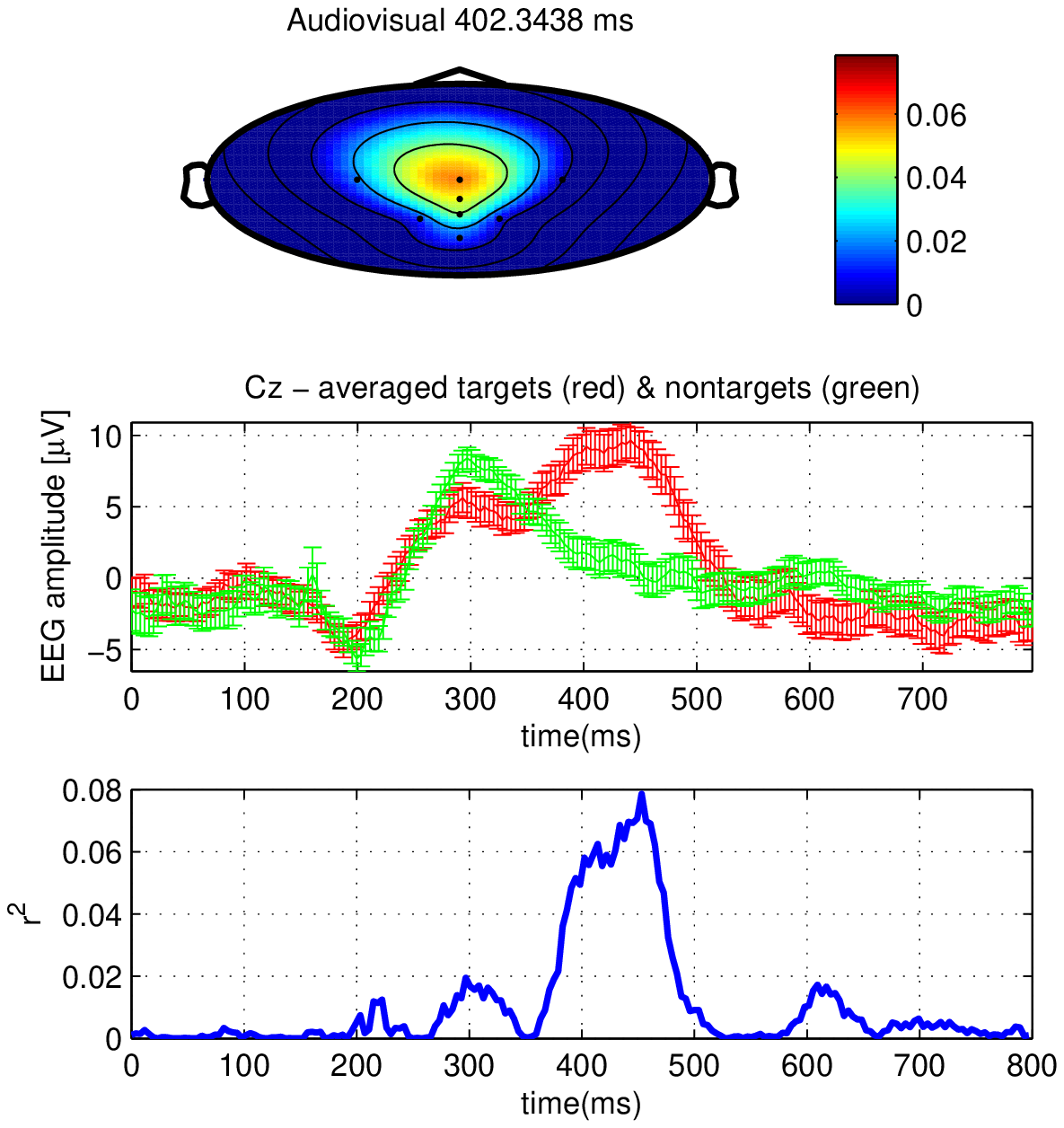}
	\caption{Audiovisual modality only offline BCI EEG responses presented as scalp mean topography for targets - in the top panel; as the mean time series with standard deviation error--bars for non-targets (green) and targets (red) in the middle panel; and as the signed statistical difference (i.e., signed $r^2$) value which evaluates the discriminability between the two types of ERPs - in the bottom panel. The figure was created with \textsf{P300GUI}~\cite{p300gui}.}\label{fig:erpAV}
\end{figure}

The collected EEG signals have been analyzed offline using \textsc{P300GUI}~\cite{p300gui} classification toolbox. The results of the analysis has been presented in Figures~\ref{fig:erpAUDIO}--\ref{fig:erpAV}. The top panels in each of the offline BCI EEG experiment results figure present the scalp topographies for targets at an ERP latency related to the maximum difference between target and non--target responses. The middle panels of the result figures visualize the differences between mean target and non--target responses as the time series. The visual and audiovisual modalities (Figures~\ref{fig:erpVISUAL}~and~\ref{fig:erpAV}, respectively) have resulted with the classical $P300$ responses with positive deflection after $300$~ms from the stimulus onset. The auditory only modality resulted with a very interesting earlier separable component (see Figure~\ref{fig:erpAUDIO}) starting around $250$~ms after the stimulus onset, which has an opposite polarity comparing the the vision related modalities. This is a very interesting result for the future development of the auditory modality only based BCI paradigms.

\section{Conclusions}

The results of the conducted psychophysical experiments have shown that the auditory modality only has the same cognitive load in the BCI paradigm application as the traditional visual or audiovisual cases. The good news is that the less explored, in BCI research, auditory spatial modality could create a perfect alternative for the subjects with limited or completely lost vision, which situation is common in case of totally--locked--in--syndrome patients.

The presented preliminary EEG responses in the same offline BCI setting confirmed also the hypothesis of the possible successful replacement of the traditional spatial visual or audiovisual paradigms with the auditory one. We also presented the very interesting and new early auditory response, which could possibly speed up and enhance the intentional response classification in auditory modality only BCI applications. 

The presented research and results are a step forward in the dynamic state--of--the--art research leading into development of less mentally demanding and with higher accuracy BCI paradigms which are very much awaited by the patients in need.

We plan the following series of experiments with larger number of subjects to further test and optimize the proposed spatial auditory speller BCI paradigm, which will also include a larger number of Japanese letters to be spelled. 

\section*{Acknowledgments}
This research was supported in part by the Strategic Information and Communications R\&D Promotion Programme no. 121803027 of The Ministry of Internal Affairs and Communication in Japan, and by KAKENHI, the Japan Society for the Promotion of Science grant no. 12010738.

\bibliography{abci}

\newpage
\section*{Figure Legends}

\begin{figure}[H]
	\centering
	\includegraphics[width=0.7\linewidth]{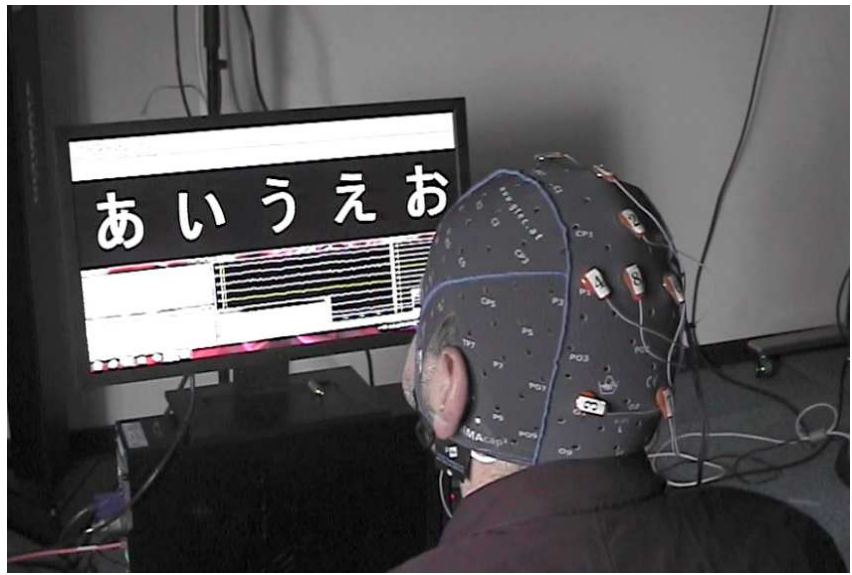}
	\vspace{0.05cm}
	\caption{Visual, auditory and audiovisual BCI speller test paradigm environment based on \textsf{BCI2000} software~\cite{bci2000book}. The subject in the picture wears also EEG \textsf{g.SAHARA} electrodes connected to \textsf{g.MOBIlab+} amplifier by \textsf{g.tec}. In this paper we conduct the psychophysical and the preliminary EEG, in offline BCI setting, experiments to evaluate possible differences related to various stimulus perceptions.}
\label{fig:aiueoBCI}
\end{figure}

\begin{figure}[H]
	\centering
	\includegraphics[width=0.5\linewidth,angle=-90]{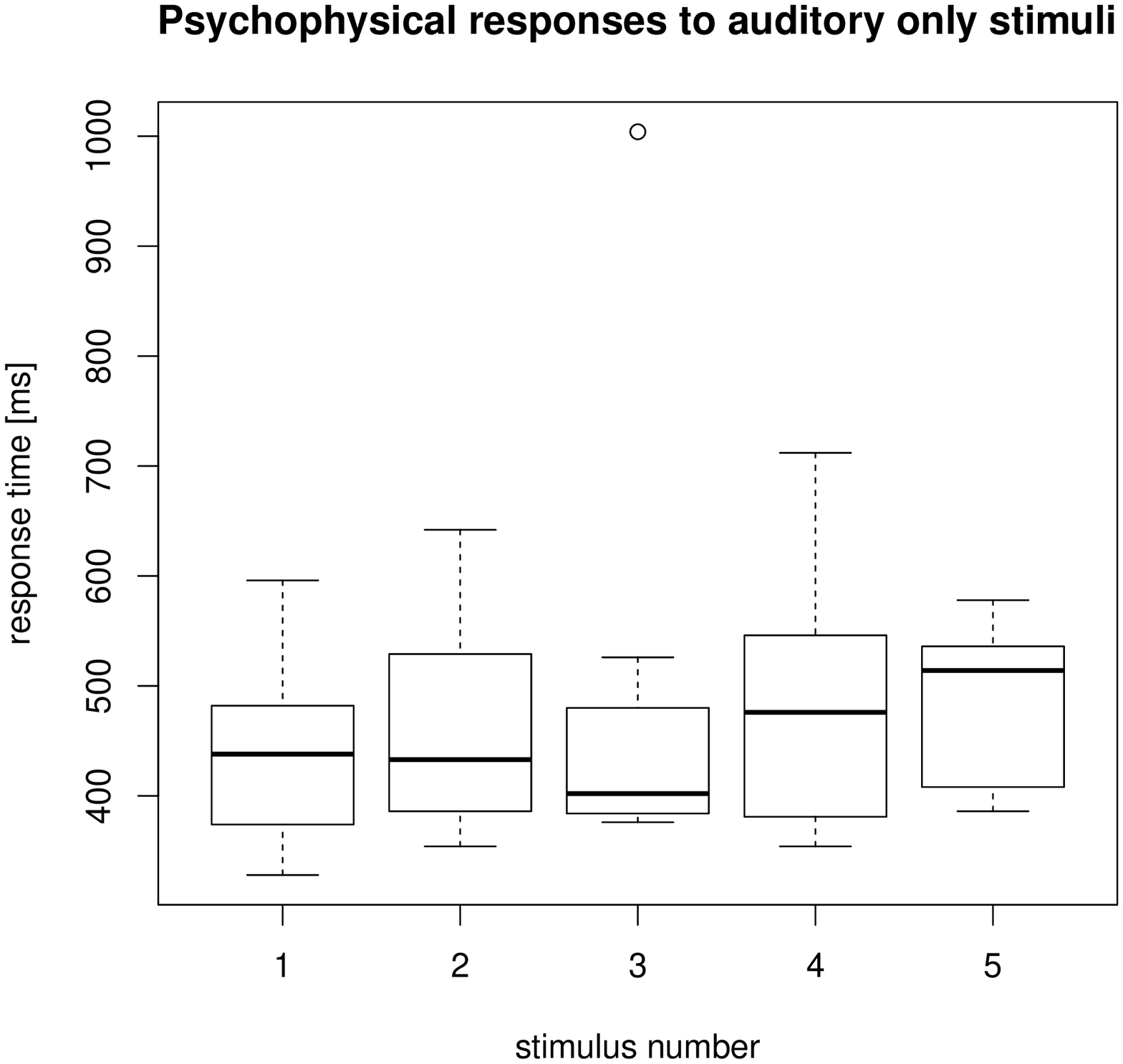}
\caption{Auditory modality only psychophysical experiment results in form of boxplots with response distributions of the five different stimuli. The differences in response times among stimuli directions are not significant.}
\label{fig:audio}
\end{figure}

\begin{figure}[H]
	\centering
	\includegraphics[width=0.5\linewidth,angle=-90]{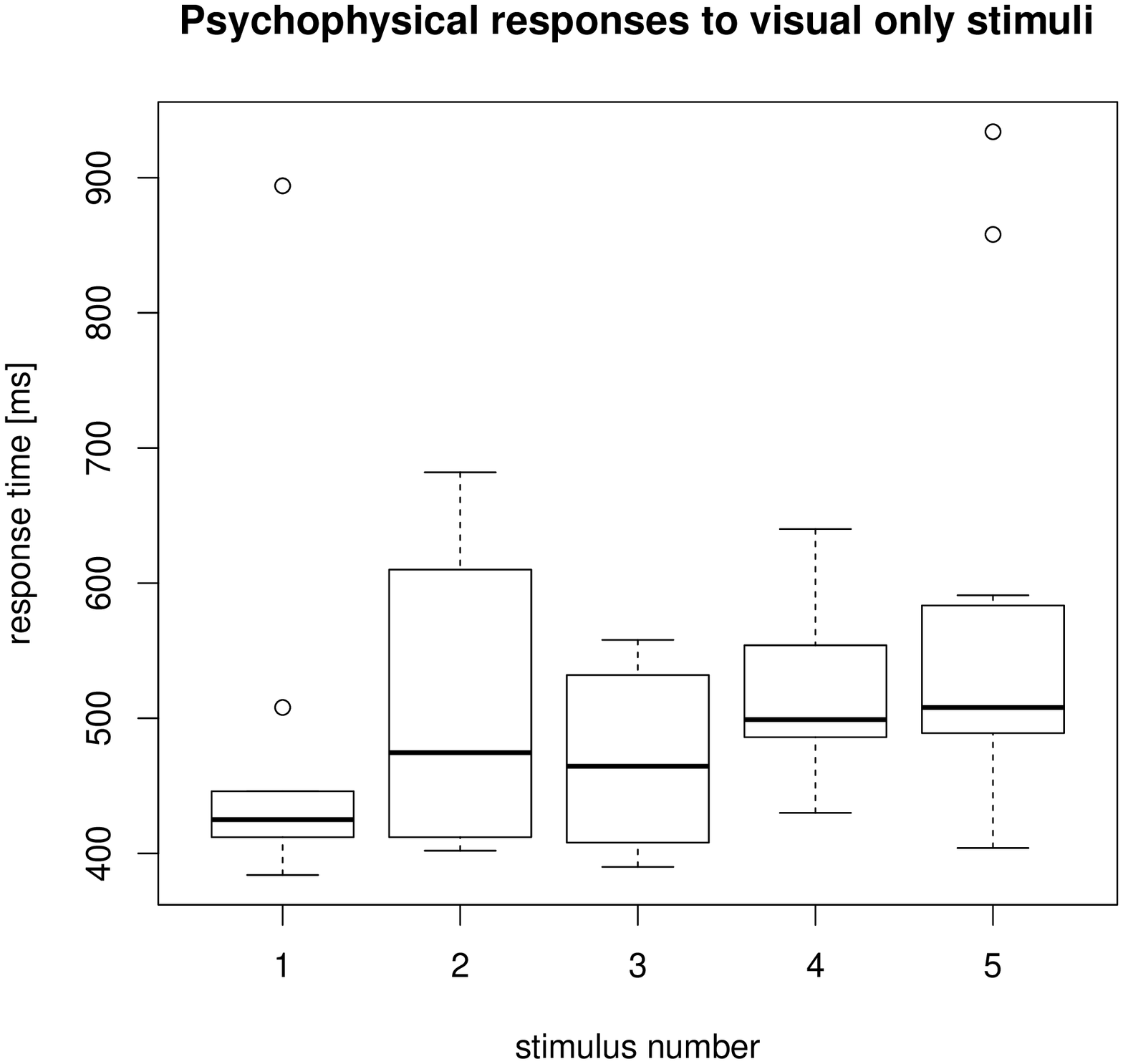}
	\caption{Visual modality only psychophysical experiment results in form of boxplots with response distributions of the five different stimuli. The differences in response times among stimuli directions are not significant.}\label{fig:video}
\end{figure}

\begin{figure}[H]
	\centering
	\includegraphics[width=0.5\linewidth,angle=-90]{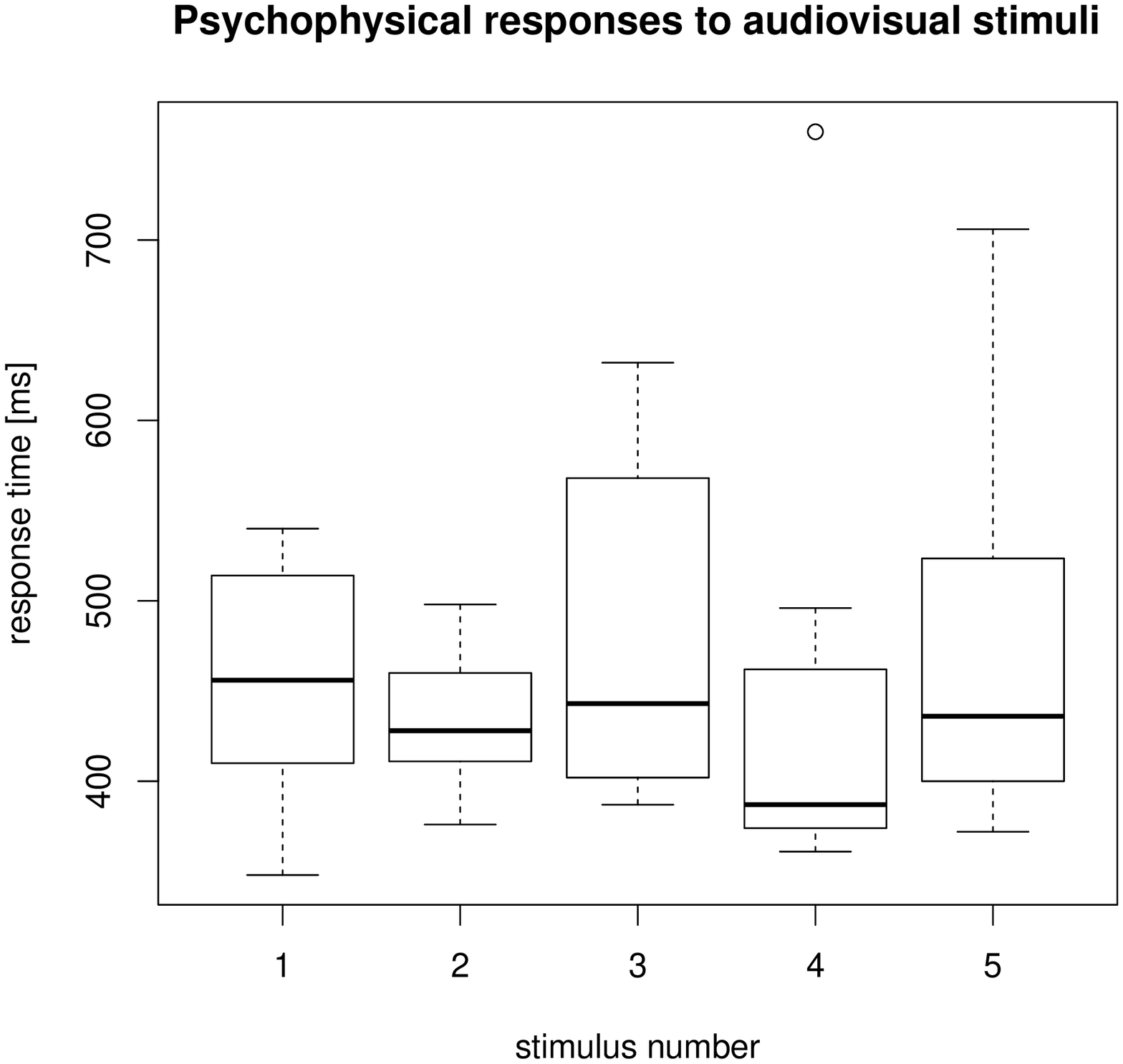}
	\caption{Audiovisual modality psychophysical experiment results in form of boxplots with response distributions of the five different stimuli. The differences in response times among stimuli directions are not significant.}\label{fig:audiovideo}
\end{figure}

\begin{figure}[H]
	\centering
	\includegraphics[width=0.5\linewidth,angle=-90]{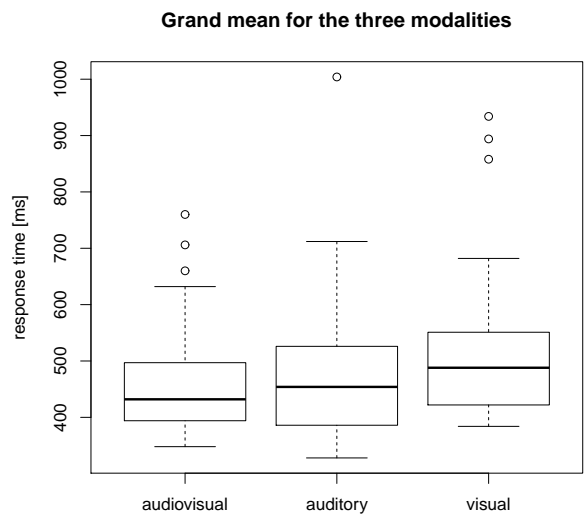}
	\caption{Grand mean for all modality psychophysical experiments as in Figures~\ref{fig:audio}--\ref{fig:audiovideo}. The median response time delays to the proposed auditory only modality are between audiovisual (the fastest) and visual (the slowest) results.}
\label{fig:multimodal}
\end{figure}

\begin{description}
	\item[Figure~\ref{fig:aiueoBCI}] Visual, auditory and audiovisual BCI speller test paradigm environment based on \textsf{BCI2000} software~\cite{bci2000book}. The subject in the picture wears also EEG \textsf{g.SAHARA} electrodes connected to \textsf{g.MOBIlab+} amplifier by \textsf{g.tec}. In this paper we conduct the psychophysical and the preliminary EEG, in offline BCI setting, experiments to evaluate possible differences related to various stimulus perceptions.
	\item[Figure~\ref{fig:audio}] Auditory modality only psychophysical experiment results in form of boxplots with response distributions of the five different stimuli. The differences in response times among stimuli directions are not significant.
	\item[Figure~\ref{fig:video}] Visual modality only psychophysical experiment results in form of boxplots with response distributions of the five different stimuli. The differences in response times among stimuli directions are not significant.
	\item[Figure~\ref{fig:audiovideo}] Audiovisual modality psychophysical experiment results in form of boxplots with response distributions of the five different stimuli. The differences in response times among stimuli directions are not significant.
	\item[Figure~\ref{fig:multimodal}] Grand mean for all modality psychophysical experiments as in Figures~\ref{fig:audio}--\ref{fig:audiovideo}. The median response time delays to the proposed auditory only modality are between audiovisual (the fastest) and visual (the slowest) results.
	\item[Figure~\ref{fig:erpAUDIO}] Auditory modality only offline BCI EEG responses presented as scalp mean topography for targets - in the top panel; as the mean time series with standard deviation error--bars for non-targets (green) and targets (red) in the middle panel; and as the signed statistical difference (i.e., signed $r^2$) value which evaluates the discriminability between the two types of ERPs - in the bottom panel. The figure was created with \textsf{P300GUI}~\cite{p300gui}.
	\item[Figure~\ref{fig:erpVISUAL}] Visual modality only offline BCI EEG responses presented as scalp mean topography for targets - in the top panel; as the mean time series with standard deviation error--bars for non-targets (green) and targets (red) - in the middle panel; and as the signed statistical difference (i.e., signed $r^2$) value which evaluates the discriminability between the two types of ERPs - in the bottom panel. The figure was created with \textsf{P300GUI}~\cite{p300gui}.
	\item[Figure~\ref{fig:erpAV}] Audiovisual modality only offline BCI EEG responses presented as scalp mean topography for targets - in the top panel; as the mean time series with standard deviation error--bars for non-targets (green) and targets (red) in the middle panel; and as the signed statistical difference (i.e., signed $r^2$) value which evaluates the discriminability between the two types of ERPs - in the bottom panel. The figure was created with \textsf{P300GUI}~\cite{p300gui}.
\end{description}


\newpage
\section*{Figures}

\end{document}